\documentclass[aip,apl,reprint,groupedaddress,preprintnumber,twocolumn,showpacs]{revtex4-1}
\usepackage{graphicx}
\usepackage{color}
\usepackage{dcolumn}

\begin{document}


\title{Large thermopower in the antiferromagnetic semiconductor BaMn$_2$Bi$_2$}
\author{Kefeng Wang}
\affiliation{Condensed Matter Physics and Materials Science Department, Brookhaven National Laboratory, Upton New York 11973 USA}
\author{C. Petrovic}
\affiliation{Condensed Matter Physics and Materials Science Department, Brookhaven National Laboratory, Upton New York 11973 USA}

\date{\today}

\begin{abstract}
We report electrical and thermal transport properties of Mn-based material BaMn$_2$Bi$_2$ with ThCr$_2$Si$_2$ structure. The resistivity of the antiferromagnetic BaMn$_2$Bi$_2$ shows a metal-semiconductor transition at $\sim 80$ K with decreasing temperature. Correspondingly, the thermopower $S$ shows a peak at the same temperature, approaching ~150 $\mu$V/K.  With increasing temperature $S$ decreases to about 125 $\mu$V/K at the room temperature. The magnetic field enhances the peak value to 210 $\mu$V/K. The Hall resistivity reveals an abrupt change of the carrier density close to the metal-semiconductor transition temperature.

\end{abstract}

\maketitle


Thermoelectric materials with high Seebeck coefficient $S$ (thermopower) have been attracting significant attention because of potential applications, particularly in waste heat recovery.\cite{TE1,TE2,TE3,TE4} High figure of merit (ZT=$\sigma S^2T/\kappa$, where $\sigma$ and $\kappa$ are the electrical and thermal conductivity, respectively) usually requires high thermopower. This raises considerable interest in exploratory synthesis of strongly correlated electron materials. Thermopower represents an electrical current entropy flow and therefore the charge/spin/orbital degrees of freedom might be manipulated for its enhancement, particularly around metal-insulator transitions.\cite{SCE1,SCE2,SCE3,SCE4} For example, giant thermopower and a record high thermoelectric power factor up to $S^2/\rho\sim 2300~\mu WK^{-2}cm^{-1}$ was observed in FeSb$_2$ with narrow energy gaps and correlated bands.\cite{fesb,fesb1,fesb2,fesb3}

Since the discovery of high temperature superconductivity in layered iron pnictide and iron chalcogenide compounds, the large diversity of the layered transitional metal pnictide compounds have been explored.\cite{iron1,iron2} In particular, doped $AM_2Pn_2$ ($A$=Ca, Sr, Ba or Eu, $M$=Fe, Mn, Rh or Co,  and $Pn$ is pnictide or chalcogenide element) with ThCr$_2$Si$_2$ (122-type) structure have been thoroughly investigated. Besides Fe-based high temperature superconductivity, high thermopower with metallic conduction was observed.\cite{Fe1,Fe2,Fe3} Mn-based $AM_2Pn_2$ materials usually exhibit magnetic ground states with strong correlations.\cite{BaMn2As2-1,BaMn2As2-2,LaMnAsO} BaMn$_2$As$_2$ and BaMn$_2$Sb$_2$ are antiferromagnetic semiconductors due to the strong Hund's coupling and the stability of the half-filled $d$-shell of the Mn$^{+2}$ ions.\cite{BaMn2As2-1,BaMn2As2-2,BaMn2Sb2-1,BaMn2Sb2-2} Both were predicted to exhibit large Seebeck coefficient.\cite{BaMn2As2-2,BaMn2Sb2-1} Therefore, it is of interest to explore thermoelectric properties of isostructural and semiconducting BaMn$_2$Bi$_2$.\cite{BaMn2Bi2}

Here we report electrical and thermal transport properties of Mn-based material BaMn$_2$Bi$_2$ with ThCr$_2$Si$_2$ structure. The resistivity of the antiferromagnetic BaMn$_2$Bi$_2$ shows a metal-semiconductor transition at $\sim 80$ K with decreasing temperature. Correspondingly, the thermopower $S$ shows a peak at the same temperature and the value approaches 150 $\mu$V/K.  With increasing temperature $S$ decreases, but is still about 120 $\mu$V/K at the room temperature. The magnetic field enhances the peak value to 210 $\mu$V/K. The Hall resistivity reveals an abrupt change of the carrier density close to the metal-semiconductor transition temperature.


Single crystals of BaMn$_2$Bi$_2$ were grown using a high-temperature self-flux method.\cite{Fisk,Canfield} X-ray diffraction (XRD) data were taken with Cu K$_{\alpha}$ ($\lambda=0.15418$ nm) radiation of Rigaku Miniflex powder diffractometer. Electrical transport measurements up to 9 T were conducted in Quantum Design PPMS-9 with conventional four-wire method. In the in-plane measurements, the current path was in the \textit{ab}-plane, whereas magnetic field was parallel to the \textit{c}-axis. Thermal transport properties were measured in Quantum Design PPMS-9 from 2 K to 350 K using one-heater-two-thermometer method. The direction of heat and electric current transport was along the $ab$-plane of single grain crystals with magnetic field along the \textit{c}-axis and perpendicular to the heat/electrical current. The relative error in our measurement was $\frac{\Delta \kappa}{\kappa}\sim$5$\%$ and $\frac{\Delta S}{S}\sim$5$\%$ based on Ni standard measured under identical conditions.


Fig. 1(a) shows the powder XRD pattern of flux grown BaMn$_2$Bi$_2$ crystals, which were fitted by RIETICA software.\cite{rietica} All reflections can be indexed in the I4/mmm space group, and the crystal structure features polyanionic [Mn$_2$Bi$_2$]$^{2-}$ layers separated by Ba ions (Fig. 1(b)). The crystals are plate-like and the base-plane is $ab$-plane (inset in Fig. 1(a)). The temperature dependence of the magnetization is shown in Fig. 1(c). The high anisotropy and decreasing magnetization with temperature suggest collinear antiferromagnetic order below room temperature, in agreement with previous result.\cite{BaMn2Bi2}

\begin{figure}[tbp]
\includegraphics[scale=0.42]{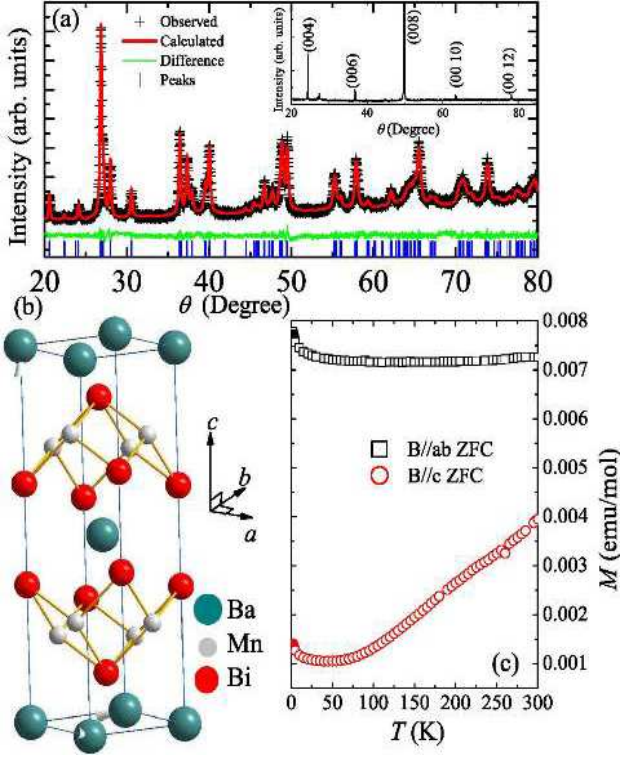}
\caption{(a) Powder XRD patterns and structural refinement results. The data were shown by ($+$) , and the fit is given by the red solid line. The difference curve (the green solid line) is offset. The inset is the XRD pattern of single crystal showing the base $ab$-plane. (b) The crystal structure of BaMn$_2$Bi$_2$. Ba ions are denoted by the largest balls, while Bi and Mn atoms are denoted as the medium and smallest balls respectively. (c) The magnetization of BaMn$_2$Bi$_2$ crystal as function of temperature with 1 T magnetic field parallel to $ab$-plane and $c$-axis respectively.}
\end{figure}

Fig. 2 shows the temperature dependence of the electric resistivity in the $ab$-plane $\rho$, thermopower $S$ and thermal conductivity $\kappa$. The resistivity decreases with increasing temperature showing semiconducting behavior up to $\sim 70$ K. At higher temperatures $\rho(T)$ is metallic (Fig. 2(a)). The thermopower $S$ is 125 $\mu$ V/K at 350 K and remains nearly constant with decreasing temperature down to 150 K. The $S$ shows a peak at $\sim 75$ K and the peak value is 150 $\mu$V/K. Below that temperature $S$ decreases to zero gradually with decreasing temperature to 2 K (Fig. 2(b)). The thermal conductivity $\kappa$ shows a phonon peak at about 30 K and the maximum value is about $10$ W/K m (Fig. 2(c)). The 9 T magnetic field has small influence on the electric resistivity and thermal conductivity, but enhances the peak value of the thermopower to 210 $\mu$V/K. The thermopower of BaMn$_2$Bi$_2$ is close to the value of typical thermoelectric materials such as PbTe and Bi$_2$Se$_3$, and the thermal conductivity is also small. However, the maximum value of ZT is $\sim0.005$ at 300 K due to high electric resistivity. It is reported that K-doping in Ba sites could induce the change of the ground state from semiconductor to metal with significant suppression of the resistivity.\cite{BaMn2Bi2} But the carrier doping could also decrease the Seebeck coefficient, such as the case in doped FeSb$_2$. This could compensate the suppression of the resistivity and make the enhancement of ZT smaller or even impossible. So the thermoelectric properties of K-doped BaMn$_2$Bi$_2$ deserve further study.

\begin{figure}[tbp]
\includegraphics[scale=0.4] {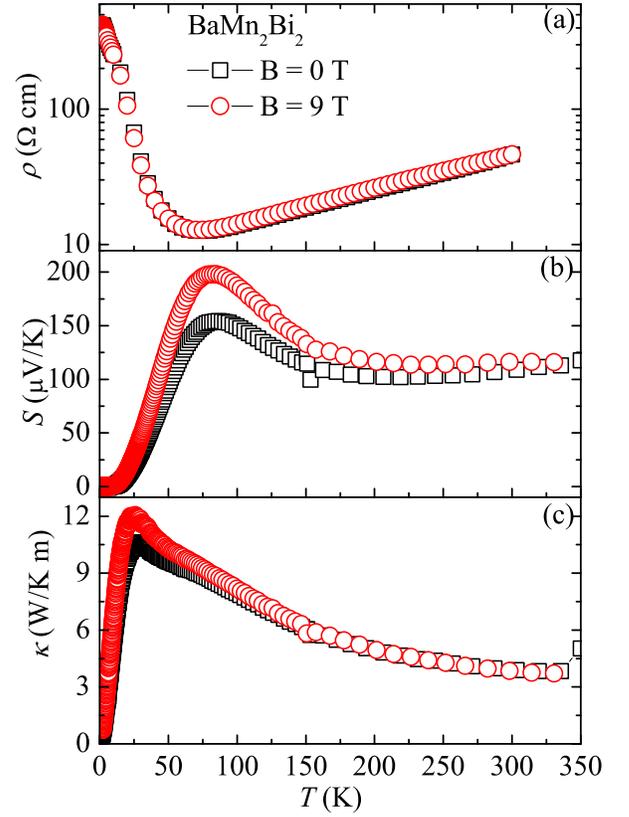}
\caption{In-plane resistivity $\rho_{ab}(T)$ (a), Seebeck coefficient $S(T)$ (b) and thermal conductivity $\kappa(T)$(c) of BaMn$_2$Bi$_2$ single crystal as a function of temperature in 0 T and 9 T magnetic field respectively.}
\end{figure}

\begin{figure}[tbp]
\includegraphics[scale=0.4] {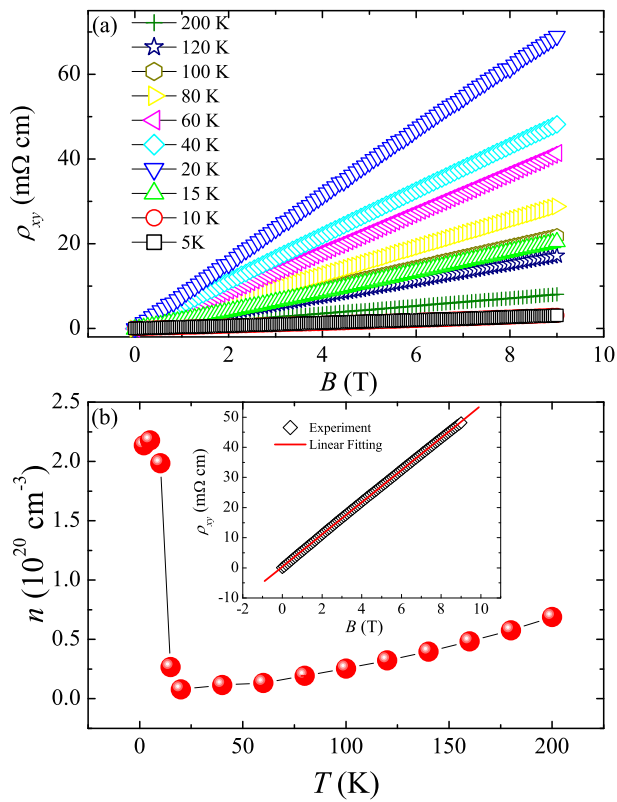}
\caption{(Color online) (a) Hall resistivity $\rho_{xy}$ as a function of the magnetic field $B$ at different temperatures. (b) The temperature dependence of the carrier density $n$ deduced from Hall resistivity. The inset shows the linear fitting of the Hall resistivity at 40 K.}
\end{figure}

Fig. 3(a) shows the Hall resistivity $\rho_{xy}$ as a function of the magnetic field $B$ at different temperatures from 5 K to 200 K. In all temperature, the Hall resistivity is positive. This indicates the hole carriers in BaMn$_2$Bi$_2$, consistent with the positive Seebeck coefficient in Fig. 2(b). Besides, the Hall resistivity shows linear field dependence and indicates single-band conduction. In single band semiconductor, the Hall resistivity can be described by single-band Hall resistivity $\rho_{xy}=\frac{B}{n|e|}$ where $n$ is the carrier density and $e$ is the electron charge. The Hall resistivity of BaMn$_2$Bi$_2$ can be fitted very well by this formula (the red line in the inset of Fig. 3(b)) and the temperature dependence of the carrier density $n$ derived from the linear fitting is shown in Fig. 3(b).

With increase in temperature, the slope of the Hall resistivity is nearly constant between 5 K and 15 K, and then increases indicating the decrease of the carrier density (Fig. 3(b)). At $\sim 40$ K, there is a large decrease in the slope of the Hall resistivity and carrier density. This position is close to the semiconductor-metal transition temperature and the peak position of Seebeck coefficient. After that, the carrier density shows a slow increase with increasing temperature.

Seebeck coefficient in a semiconductor is the sum of three different contributions: the diffusion term $S_{diff}$, the spin-dependent scattering term and the phonon-drag term $S_{drag}$ due to electron-phonon coupling. The diffusion term of a single-band metal always shows linear temperature dependence and the non-monotonic behavior can only come from spin scattering or phonon-drag.\cite{TEP1,TEP2} In BaMn$_2$Bi$_2$, the magnetic transition temperature (above 400 K) is much higher than the peak position of the Seebeck coefficient and the Seebeck coefficient does not show significant magnetic field dependence expect around the metal-semiconductor transition. This indicates the spin-dependent scattering should not dominate the Seebeck coefficient in BaMn$_2$Bi$_2$. The contribution of phonon-drag term gives $\sim T^3$ dependence for $T<<\Theta_D$, $\sim 1/T$ for $T\geq\Theta_D$ (where $\Theta_D$ is the Debye Temperature), and a peak structure for $\sim\frac{\Theta_D}{5}$.\cite{TEP2,TEP2} The Debye temperature of BaMn$_2$Bi$_2$ is $\sim 150$ K.\cite{BaMn2Bi2} The peak structure from phonon-drag should be at $\sim 30$ K which is rather different from the observed peak position ($\sim 75$ K) in Fig. 2(b). Hence, the peak of Seebeck coefficient in BaMn$_2$Bi$_2$ should not come from the spin-dependent scattering term and the phonon-drag term. Instead, its origin is in the sharp change in the carrier density and the metal-semiconductor transition which is related the strongly correlated effect.

In summary, we report the electronic and thermal transport properties of Mn-based material BaMn$_2$Bi$_2$ with ThCr$_2$Si$_2$ structure. Thermopower of the antiferromagnetic BaMn$_2$Bi$_2$ shows a peak at the temperature of the metal-semiconductor transition ($\sim 80$ K) of 150 $\mu$V/K.  With increasing temperature $S$ decreases slightly and the value is have 120 $\mu$V/K at room temperature. The magnetic field enhances the peak value to 210 $\mu$V/K. The Hall resistivity reveals an abrupt change of the carrier density close to the metal-semiconductor transition temperature.

\begin{acknowledgments}
We than John Warren for help with SEM measurements. Work at Brookhaven is supported by the U.S. DOE under contract No. DE-AC02-98CH10886.
\end{acknowledgments}


\end{document}